\documentclass[runningheads]{llncs} 
\usepackage[misc]{ifsym}
\usepackage{graphicx}

\usepackage[colorlinks=true,allcolors=blue]{hyperref}
\usepackage{amsmath,amssymb,amsfonts}

\usepackage{multirow}
\usepackage{booktabs}
\usepackage{placeins}
\usepackage{bm}
\usepackage{subfigure} 
\usepackage{floatrow}
\newfloatcommand{capbtabbox}{table}[][\FBwidth]
\begin{document}
\title{DRMC: A Generalist Model with Dynamic Routing for Multi-Center PET Image Synthesis
% \thanks{Supported by organization x.}
}
\titlerunning{DRMC: A Generalist Model for Multi-Center PET Image Synthesis}
\author{
Zhiwen Yang\inst{1} \and
Yang Zhou\inst{1} \and
Hui Zhang\inst{2} \and 
Bingzheng Wei\inst{3} \and 
Yubo Fan \inst{1} \and 
% Yan Xu \inst{1} \textsuperscript{\Letter} 
Yan Xu\inst{1}$^{(\textrm{\Letter},\thanks{Corresponding author})}$
} 

% \author{
% First Author\inst{1}\orcidID{0000-1111-2222-3333} \and
% Third Author\inst{3}\orcidID{2222--3333-4444-5555}
% }

%
\authorrunning{Yang et al.}
\institute{
School of Biological Science and Medical Engineering, State Key Laboratory of Software Development Environment, Key Laboratory of Biomechanics and Mechanobiology of Ministry of Education, Beijing Advanced Innovation Center for Biomedical Engineering, Beihang University, Beijing 100191, China
\\
\email{xuyan04@gmail.com} \and 
Department of Biomedical Engineering, Tsinghua University, Beijing 100084, China \and 
Xiaomi Corporation, Beijing 100085, China
% Princeton University, Princeton NJ 08544, USA \and
}
\maketitle              % typeset the header of the contribution
%
% \begin{abstract}
% We focus on developing a generalist model for multi-center positron emission tomography (PET) image synthesis. The generalizability of existing methods can still be suboptimal for a multi-center study due to domain shifts, which result from non-identical data distribution among centers with different imaging systems/protocols. While some approaches address domain shift by training specialized models for each center, they conflict with our goal of a generalist model. Compared with specialized models, a generalist model shares architecture and parameters across centers. However, we find the generalist model can suffer from the center interference issue, \textit{i.e.} the gradient directions of different centers can be inconsistent or even opposite owing to the non-identical data distribution. To mitigate such interference, we introduce a novel dynamic routing strategy with cross-layer connection that routes data from different centers to different experts. Experiments show that our proposed generalist model with dynamic routing (DRMC) exhibits the best generalizability on both known and unknown centers.
% \keywords{Multi-Center \and Positron Emission Tomography \and Synthesis \and  Generalist Model \and Dynamic Routing.}
% \end{abstract} 

\begin{abstract}
Multi-center positron emission tomography (PET) image synthesis aims at recovering low-dose PET images from multiple different centers. The generalizability of existing methods can still be suboptimal for a multi-center study due to domain shifts, which result from non-identical data distribution among centers with different imaging systems/protocols. While some approaches address domain shifts by training specialized models for each center, they are parameter inefficient and do not well exploit the shared knowledge across centers. To address this, we develop a generalist model that shares architecture and parameters across centers to utilize the shared knowledge. However, the generalist model can suffer from the center interference issue, \textit{i.e.} the gradient directions of different centers can be inconsistent or even opposite owing to the non-identical data distribution. To mitigate such interference, we introduce a novel dynamic routing strategy with cross-layer connections that routes data from different centers to different experts. Experiments show that our generalist model with dynamic routing (DRMC) exhibits excellent generalizability across centers. Code and data are available at: \href{https://github.com/Yaziwel/Multi-Center-PET-Image-Synthesis}{https://github.com/Yaziwel/Multi-Center-PET-Image-Synthesis}.
\end{abstract}

\section{Introduction} 
\label{sec:intro}
Positron emission tomography (PET) image synthesis \cite{3D-cGAN,auto-context-2017,CycleGAN-mia-2020,SGSGAN-2022,adaptive-rectification-PET-2022,external-validation-2021,FTL-PET-2022,3D-Transformer-GAN-2021,spach-Transformer-2022,3DCVT-GAN} aims at recovering high-quality full-dose PET images from low-dose ones. Despite great success, most algorithms \cite{3D-cGAN,auto-context-2017,SGSGAN-2022,adaptive-rectification-PET-2022,3D-Transformer-GAN-2021,spach-Transformer-2022,3DCVT-GAN} are specialized for PET data from a single center with a fixed imaging system/protocol. This poses a significant problem for practical applications, which are not usually restricted to any one of the centers. Towards filling this gap, in this paper, we focus on multi-center PET image synthesis, aiming at processing data from multiple different centers.

However, the generalizability of existing models can still be suboptimal for a multi-center study due to domain shift, which results from non-identical data distribution among centers with different imaging systems/protocols (see Fig.~\ref{fig:center_interference} (a)). Though some studies have shown that a specialized model (\textit{i.e.} a convolutional neural network (CNN) \cite{CycleGAN-mia-2020,external-validation-2021} or Transformer \cite{spach-Transformer-2022} trained on a single center) exhibits certain robustness to different tracer types \cite{spach-Transformer-2022}, different tracer doses \cite{CycleGAN-mia-2020}, or even different centers \cite{external-validation-2021}, such generalizability of a center-specific knowledge is only applicable to small domain shifts. It will suffer a severe performance drop when exposed to new centers with large domain shifts \cite{FL-MRCM}. There are also some federated learning (FL) based \cite{FL-MR-FedAVG,FL-MRCM,FTL-PET-2022} medical image synthesis methods that improve generalizability by collaboratively learning a shared global model across centers. Especially, federated transfer learning (FTL) \cite{FTL-PET-2022} first successfully applies FL to PET image synthesis in a multiple-dose setting. Since the resultant shared model of the basic FL method \cite{FL-MR-FedAVG} ignores center specificity and thus cannot handle centers with large domain shifts, FTL addresses this by finetuning the shared model for each center/dose. However, FTL only focuses on different doses and does not really address the multi-center problem. Furthermore, it still requires a specialized model for each center/dose, which ignores potentially transferable shared knowledge across centers and scales up the overall model size.

% causes the overall model size to linearly increase with the number of centers without. 

% A recent trend, known as generalist models, is to request that a single unified model works for multiple tasks/domains, and even express generalizability to novel tasks/domains. Some pioneers \cite{moe-2017,domain_attention-2019,uniperceiver-moe-2022} have made attempts to establish generalist models by modeling various tasks/domains with shared architecture and parameters. These generalist models realize competitive performance on various high-level vision tasks like classification \cite{moe-2017,uniperceiver-moe-2022}, object detection \cite{domain_attention-2019}, \textit{etc.} 
A recent trend, known as generalist models, is to request that a single unified model works for multiple tasks/domains, and even express generalizability to novel tasks/domains. By sharing architecture and parameters, generalist models can better utilize shared transferable knowledge across tasks/domains. Some pioneers \cite{moe-2017,domain_attention-2019,uniperceiver-2021,uniperceiver-moe-2022,ofa-ICML-2022} have realized competitive performance on various high-level vision tasks like classification \cite{moe-2017,uniperceiver-moe-2022}, object detection \cite{domain_attention-2019}, \textit{etc.}

Nonetheless, recent studies \cite{gradient-surgery-2020,uniperceiver-moe-2022} report that conventional generalist \cite{uniperceiver-2021} models may suffer from the interference issue, \textit{i.e.} different tasks with shared parameters potentially conflict with each other in the update directions of the gradient. Specific to PET image synthesis, due to the non-identical data distribution across centers, we also observe the \textbf{center interference issue} that the gradient directions of different centers may be inconsistent or even opposite (see Fig.~\ref{fig:center_interference}). This will lead to an uncertain update direction that deviates from the optimal, resulting in sub-optimal performance of the model. To address the interference issue, recent generalist models \cite{domain_attention-2019,uniperceiver-moe-2022} have introduced dynamic routing \cite{DynamicNetwork-survey-2021} which learns to activate experts (\textit{i.e.} sub-networks) dynamically. The input feature will be routed to different selected experts accordingly so as to avoid interference. Meanwhile, different inputs can share some experts, thus maintaining collaboration across domains. In the inference time, the model can reasonably generalize to different domains, even unknown domains, by utilizing the knowledge of existing experts. In spite of great success, the study of generalist models rarely targets the problem of multi-center PET image synthesis. 

In this paper, inspired by the aforementioned studies, we innovatively propose a generalist model with \textbf{D}ynamic \textbf{R}outing for \textbf{M}ulti-\textbf{C}enter PET image synthesis, termed DRMC. To mitigate the center interference issue, we propose a novel dynamic routing strategy to route data from different centers to different experts. Compared with existing routing strategies, our strategy makes an improvement by building cross-layer connections for more accurate expert decisions. Extensive experiments show that DRMC achieves the best generalizability on both known and unknown centers. Our contribution can be summarized as:

\begin{itemize} 
\item A generalist model called DRMC is proposed, which enables multi-center PET image synthesis with a single unified model.
\item A novel dynamic routing strategy with cross-layer connection is proposed to address the center interference issue. It is realized by dynamically routing data from different centers to different experts.
\item Extensive experiments show that DRMC exhibits excellent generalizability over multiple different centers.

\end{itemize}

\section{Method} 
% \subsection{Preliminaries} We first introduce a quantification metric to facilitate the analysis of center interference. Following the paper, we estimate the change in loss $\mathcal{L}_i$ of the $i$-th center task, when optimizing the shared parameters $\theta$ according to the $j$-th center task $\mathcal{L}_j$ as:
% \begin{equation} 
% \begin{align}
% \Delta_j \mathcal{L}_{i}\left(X_i\right) &\doteq \mathbb{E}_{X_j}\left(\mathcal{L}_{i}\left(X_i ; \theta\right)-\mathcal{L}_{i}(X_i ; \theta-\lambda \frac{\nabla_\theta \mathcal{L}_{j}\left(X_j\right)}{\left\|\nabla_\theta \mathcal{L}_{j}\left(X_j\right)\right\|})\right), \\
% &\approx \lambda \mathbb{E}_{X_j}\left({\frac{\nabla_\theta \mathcal{L}_{j}\left(X_j\right)}{\left\|\nabla_\theta \mathcal{L}_{j}\left(X_j\right)\right\|}}^T \nabla_\theta \mathcal{L}_{i}\left(X_i\right)\right), 
% \end{align}
% \end{equation} 
% where $X_i$ and $X_j$ are the sampled training batches of the $i$-th and $j$-th centers, respectively. Then, the interference of the $j$-th center task on the $i$-th center task can be quantified as:
% \begin{equation}
% \mathcal{I}_{i, j}=\mathbb{E}_{X_i}\left(\frac{\Delta_j \mathcal{L}_{i}\left(X_i\right)}{\Delta_i \mathcal{L}_{i}\left(X_i\right)}\right),
% \end{equation}
% where the denominator is used to normalize the loss change scale. 
\subsection{Center Interference Issue} 
\label{subsec:interference}
\begin{figure*}[t]
\centering
\subfigure[]{
    \includegraphics[height=1.2in]{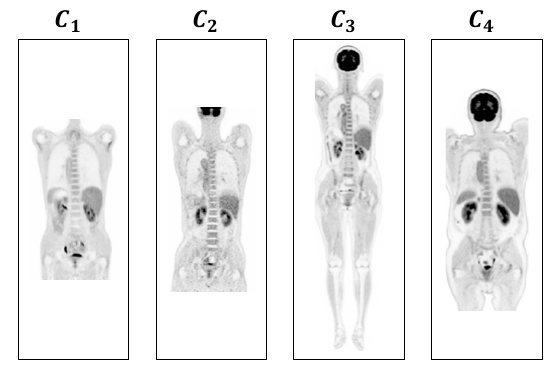}

}
\subfigure[]{
    \includegraphics[height=1.2in]{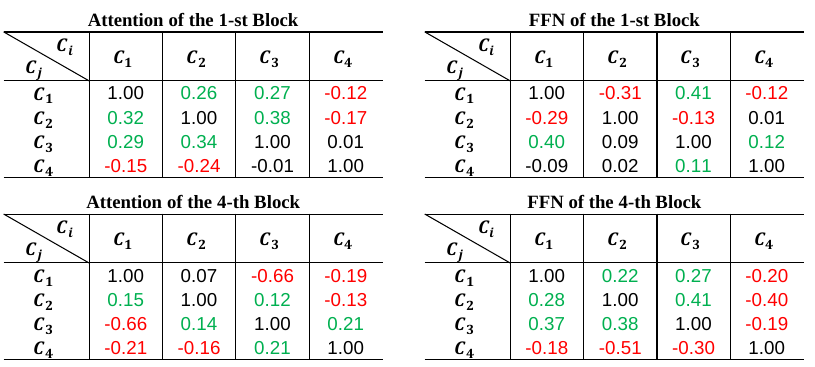}

}

\caption{(a) Examples of PET images at different Centers. There are domain shifts between centers. (b) The interference metric $\mathcal{I}_{i,j}$ \cite{uniperceiver-moe-2022} of the center $C_j$ on the center $C_i$ at the 1-st/4-th blocks as examples. The red value indicates that $C_j$ has a negative impact on $C_i$, and the green value indicates that $C_j$ has a positive impact on $C_i$.}
\label{fig:center_interference}
\end{figure*}

Due to the non-identical data distribution across centers, different centers with shared parameters may conflict with each other in the optimization process. To verify this hypothesis, we train a baseline Transformer with 15 base blocks (Fig.~\ref{fig:overall_arch} (b)) over four centers. Following the paper \cite{uniperceiver-moe-2022}, we calculate the gradient direction interference metric $\mathcal{I}_{i, j}$ of
the $j$-th center $C_j$ on the $i$-th center $C_i$. As shown in Fig.~\ref{fig:center_interference} (b), interference is observed between different centers at different layers. This will lead to inconsistent optimization and inevitably degrade the model performance. Details of $\mathcal{I}_{i, j}$ \cite{uniperceiver-moe-2022} are shown in the \textbf{supplement}.

\subsection{Network Architecture\label{sec:overall_arch}} 

The overall architecture of our DRMC is shown in Fig.~\ref{fig:overall_arch} (a). DRMC firstly applies a 3$\times$3$\times$3 convolutional layer for shallow feature extraction. Next, the shallow feature is fed into $N$ blocks with dynamic routing (DRBs), which are expected to handle the interference between centers and adaptively extract the deep feature with high-frequency information. The deep feature then passes through another 3$\times$3$\times$3 convolutional layer for final image synthesis. In order to alleviate the burden of feature learning and stabilize training, DRMC adopts global residual learning as suggested in the paper \cite{DnCNN} to estimate the image residual from different centers. In the subsequent subsection, we will expatiate the dynamic routing strategy as well as the design of the DRB.

\subsection{Dynamic Routing Strategy} 
\begin{figure*}[t]
	\centering
	\includegraphics[width=0.82\textwidth]{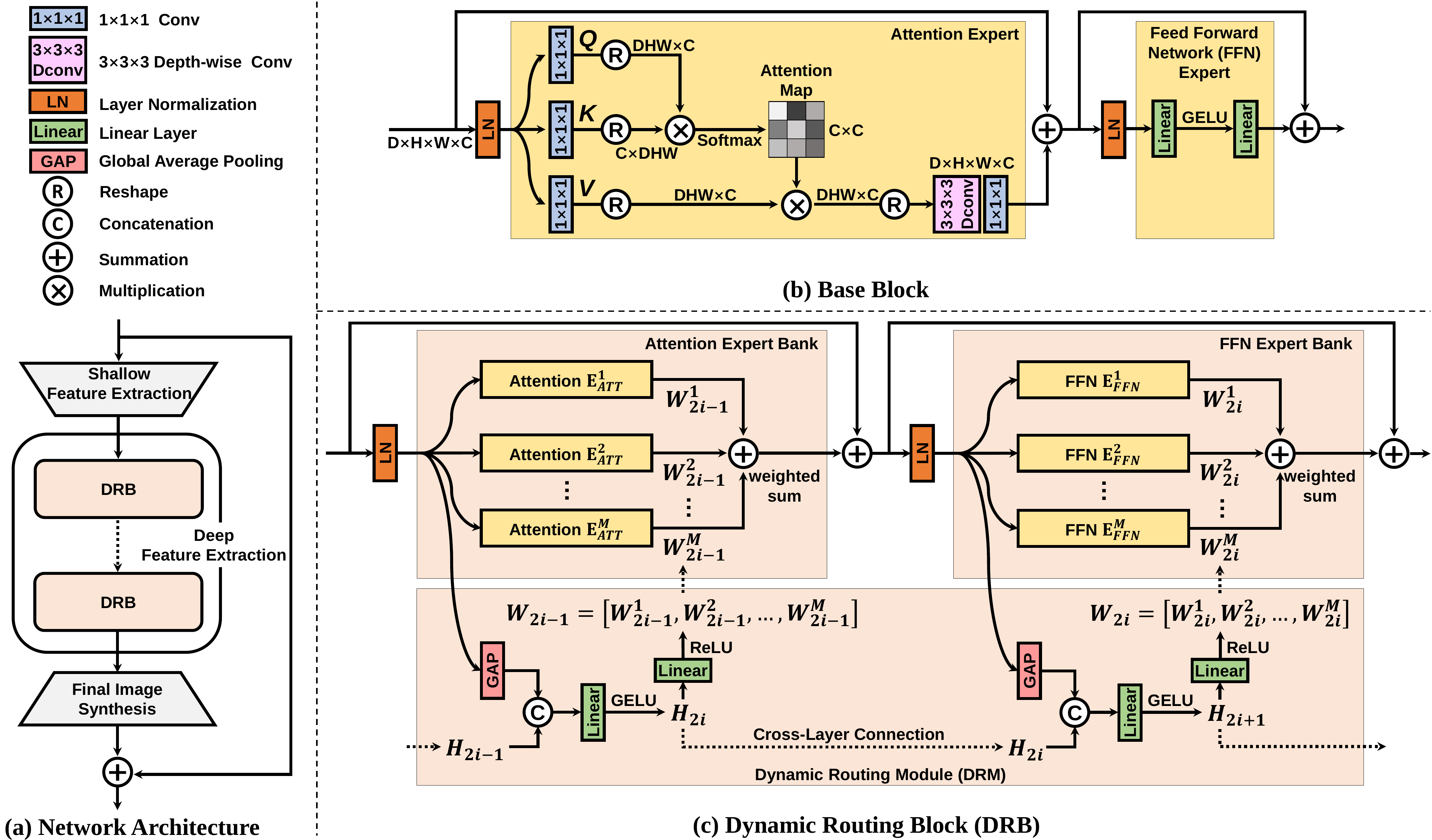}
    \caption{The framework of our proposed DRMC}
	\label{fig:overall_arch}
\end{figure*}

We aim at alleviating the center interference issue in deep feature extraction. Inspired by prior generalist models \cite{moe-2017,domain_attention-2019,uniperceiver-moe-2022}, we specifically propose a novel dynamic routing strategy for multi-center PET image synthesis. The proposed dynamic routing strategy can be flexibly adapted to various network architectures, such as CNN and Transformer. To utilize the recent advance in capturing global contexts using Transformers \cite{spach-Transformer-2022}, without loss of generality, we explore the application of the dynamic routing strategy to a Transformer block, termed dynamic routing block (DRB, see Fig.~\ref{fig:overall_arch} (c)). We will introduce our dynamic routing strategy in detail from four parts: base expert foundation, expert number scaling, expert dynamic routing, and expert sparse fusion.

 \noindent
\textbf{Base Expert Foundation.} As shown in \autoref{fig:overall_arch} (b), we first introduce an efficient base Transformer block (base block) consisting of an attention expert and a feed-forward network (FFN) expert. Both experts are for basic feature extraction and transformation. To reduce the complexity burden of the attention expert, we follow the paper \cite{spach-Transformer-2022} to perform global channel attention with linear complexity instead of spatial attention \cite{attention_all_u_need_2017}. Notably, as the global channel attention may ignore the local spatial information, we introduce depth-wise convolutions to emphasize the local context after applying attention. As for the FFN expert, we make no modifications to it compared with the standard Transformer block \cite{attention_all_u_need_2017}. It consists of a 2-layer MLP with GELU activation in between.

% \noindent
% \textbf{Expert Number Scaling.} Due to conflicts in gradient update directions of different centers on the same expert, we decide to increase the number of expert layers in the baseline block to $M$ to mitigate interference. Specifically, each Transformer block has an attention expert bank $\mathbf{E}_{ATT} = [\mathbf{E}^1_{ATT}, \mathbf{E}^2_{ATT}, ..., \mathbf{E}^M_{ATT}]$ and an FFN expert bank $\mathbf{E}_{FFN} = [\mathbf{E}^1_{FFN}, \mathbf{E}^2_{FFN}, ..., \mathbf{E}^M_{FFN}]$, both of which have $M$ experts as described in the baseline block. However, it does not mean that we prepare specific experts for each center. Although using center-specific experts can address the interference problem, it is hard for the model to reasonably generalize to new centers that did not emerge in the training stage.

\noindent
\textbf{Expert Number Scaling.} Center interference is observed on both attention experts and FFN experts at different layers (see Fig.~\ref{fig:center_interference} (b)). This indicates that a single expert can not be simply shared by all centers. Thus, we increase the number of experts in the base block to $M$ to serve as expert candidates for different centers. Specifically, each Transformer block has an attention expert bank $\mathbf{E}_{ATT} = [\mathbf{E}^1_{ATT}, \mathbf{E}^2_{ATT}, ..., \mathbf{E}^M_{ATT}]$ and an FFN expert bank $\mathbf{E}_{FFN} = [\mathbf{E}^1_{FFN}, \mathbf{E}^2_{FFN}, ..., \mathbf{E}^M_{FFN}]$, both of which have $M$ base experts. However, it does not mean that we prepare specific experts for each center. Although using center-specific experts can address the interference problem, it is hard for the model to exploit the shared knowledge across centers, and it is also difficult to generalize to new centers that did not emerge in the training stage \cite{uniperceiver-moe-2022}. To address this, we turn to different combinations of experts.

\noindent
\textbf{Expert Dynamic Routing.} Given a bank of experts, we route data from different centers to different experts so as to avoid interference. Prior generalist models \cite{moe-2017,domain_attention-2019,uniperceiver-moe-2022} in high-level vision tasks have introduced various routing strategies to weigh and select experts. Most of them are independently conditioned on the information of the current layer feature, failing to take into account the connectivity of neighboring layers. Nevertheless, PET image synthesis is a dense prediction task that requires a tight connection of adjacent layers for accurate voxel-wise intensity regression. To mitigate the potential discontinuity \cite{moe-2017}, we propose a dynamic routing module (DRM, see Fig.~\ref{fig:overall_arch} (c)) that builds cross-layer connection for expert decisions. The mechanism can be formulated as:
\begin{equation}
W=\mathbf{ReLU}(\mathbf{MLP}(\left[\mathbf{GAP}(X), H\right])),
\end{equation}
 where $X$ denotes the input; $\mathbf{GAP}(\cdot)$ represents the global average pooling operation to aggregate global context information of the current layer; $H$ is the hidden representation of the previous MLP layer. ReLU activation generates sparsity by setting the negative weight to zero. It is a more suitable gating function in comparison with the commonly used softmax activation \cite{domain_attention-2019} and top-k gating \cite{moe-2017,uniperceiver-moe-2022} in our study (see Table.~\ref{tab:ablation}). $W$ is a sparse weight used to assign weights to different experts.
 
 In short, DRM sparsely activates the model and selectively routes the input to different subsets of experts. This process maximizes collaboration and meanwhile mitigates the interference problem. On the one hand, the interference across centers can be alleviated by sparsely routing $X$ to different experts (with positive weights). The combinations of selected experts can be thoroughly different across centers if violent conflicts appear. On the other hand, experts in the same bank still cooperate with each other, allowing the network to best utilize the shared knowledge across centers. 

\noindent
\textbf{Expert Sparse Fusion.} The final output is a weighted sum of each expert's knowledge using the sparse weight $W=[W^1, W^2, ..., W^M]$ generated by DRM. Given an input feature $X$, the output $\hat{X}$ of an expert bank can be obtained as:
\begin{equation}
\hat{X}=\sum_{m=1}^M W^m \cdot \mathbf{E}^m(X),
\end{equation}
where $\mathbf{E}^m(\cdot)$ represents an operator of $\mathbf{E}^m_{ATT}(\cdot)$ or $\mathbf{E}^m_{FFN}(\cdot)$.

\begin{table}[t] 
\caption{Multi-Center PET Dataset Information} 
\centering
\resizebox{0.98\textwidth}{!}{
\begin{tabular}{cc|c|c|c|c|c|c|c|c|c|c|c}
\toprule
\multicolumn{2}{c|}{Center}                             & Institution & Type       & Lesion & System    & Tracer       & Dose   & DRF & Spacing (${mm}^3$)           & Shape                          & Train         & Test \\ \midrule
\multicolumn{1}{c|}{\multirow{4}{*}{$C_{kn}$}}  & $C_1$ & $I_1$   & Whole Body & Yes    & PolarStar m660 & $^{18}$F-FDG & 293MBq & 12  & 3.15$\times$3.15$\times$1.87 & 192$\times$192$\times$$slices$ & 20            & 10   \\
\multicolumn{1}{c|}{}                           & $C_2$ & $I_2$   & Whole Body & Yes    & PolarStar Flight & $^{18}$F-FDG & 293MBq & 4   & 3.12$\times$3.12$\times$1.75 & 192$\times$192$\times$$slices$ & 20            & 10   \\
\multicolumn{1}{c|}{}                           & $C_3$\cite{EJNMMI-2021-Ultral-low-challenge} & $I_3$   & Whole Body & Yes    & United Imaging uEXPLORER & $^{18}$F-FDG & 296MBq & 10  & 1.67$\times$1.67$\times$2.89 & 256$\times$256$\times$$slices$ & 20            & 10   \\
\multicolumn{1}{c|}{}                           & $C_4$\cite{EJNMMI-2021-Ultral-low-challenge} & $I_4$   & Whole Body & Yes    & Siemens Biograph Vision Quadra & $^{18}$F-FDG & 296MBq & 10  & 1.65$\times$1.65$\times$1.65 & 256$\times$256$\times$$slices$ & 20            & 10   \\ \hline
\multicolumn{1}{c|}{\multirow{2}{*}{$C_{ukn}$}} & $C_5$ & $I_5$   & Brain      & No     & PolarStar m660 & $^{18}$F-FDG & 293MBq & 4   & 1.18$\times$1.18$\times$1.87 & 256$\times$256$\times$$slices$ & $\textendash$ & 10   \\
\multicolumn{1}{c|}{}                           & $C_6$ & $I_6$   & Whole Body & Yes    & PolarStar m660 & $^{18}$F-FDG & 293MBq & 12  & 3.15$\times$3.15$\times$1.87 & 192$\times$192$\times$$slices$ & $\textendash$ & 10   \\ \bottomrule
\end{tabular}
}
\label{tab:dataset}
\end{table}

% \subsection{Loss Function}
% We utilize Charbonnier loss \cite{CharbonnierLoss} with hyper-parameter $\epsilon$ as $10^{-3}$ to penalize pixel-wise differences between the EPET and FPET images from each center $C_k$:
% \begin{equation}
% \mathcal{L}_{C_{k}}=\sqrt{\left\|I^{EPET}_{C_k}-I^{FPET}_{C_k}\right\|^2+\epsilon^2}.
% \end{equation}
% To properly train a model that is robust to different centers, we need to balance the tasks of different centers so as to avoid some centers showing dominance during training \cite{MTL-uncertainty-weight}. Otherwise, the model tends to overfit in those centers expressing dominance and show poor performance to other centers. Therefore, we follow the prior work \cite{MTL-uncertainty-weight} in multi-task learning to weight losses of $K$ centers:
% \begin{equation}
%         \mathcal{L}=\sum_{k=1}^{K}\left(\frac{1}{2\sigma_{k}^2}\mathcal{L}_{C_k} + log\sigma_k \right), 
% \label{eq:loss}
% \end{equation} 
% where $\sigma_k$ is a learnable parameter, $\frac{1}{2\sigma_{k}^2}$ is for penalizing the loss term, and $log\sigma_k$ is a regularization term.  

\subsection{Loss Function}
We utilize the Charbonnier loss \cite{CharbonnierLoss} with hyper-parameter $\epsilon$ as $10^{-3}$ to penalize pixel-wise differences between the full-dose ($Y$) and estimated ($\hat{Y}$) PET images:
\begin{equation}
\mathcal{L}=\sqrt{\left\|Y-\hat{Y}\right\|^2+\epsilon^2}.
\end{equation} 

\section{Experiments and Results}
\subsection{Dataset and Evaluation} 
Full-dose PET images are collected from 6 different centers ($C_1\textendash C_6$) at 6 different institutions\footnote[1]{$I_1$ and $I_5$ are Peking Union Medical College Hospital; $I_2$ is Beijing Hospital; $I_3$ is Department of Nuclear Medicine, Ruijin Hospital, Shanghai Jiao Tong University School of Medicine; $I_4$ is Department of Nuclear Medicine, University of Bern; $I_6$ is Beijing Friendship Hospital.}. The data of $C_3$ and $C_4$ \cite{EJNMMI-2021-Ultral-low-challenge} are borrowed from the Ultra-low Dose PET Imaging Challenge\footnote[2]{Challenge site: \href{https://ultra-low-dose-pet.grand-challenge.org/}{https://ultra-low-dose-pet.grand-challenge.org/}. The investigators of the challenge contributed to the design and implementation of DATA, but did not participate in analysis or writing of this paper. A complete listing of investigators can be found at:\href{https://ultra-low-dose-pet.grand-challenge.org/Description/}{https://ultra-low-dose-pet.grand-challenge.org/Description/}.}, while the data from other centers were privately collected. The key information of the whole dataset is shown in Table.~\ref{tab:dataset}. Note that $C_1\textendash C_4$ are for both training and testing. We denote them as $C_{kn}$ as these centers are known to the generalist model. $C_5$ and $C_6$ are unknown centers (denote as $C_{ukn}$) that are only for testing the model generalizability. The low-dose PET data is generated by randomly selecting a certain portion of the raw scans according to the dose reduction factor (DRF), \textit{e.g.} the portion is 25$\%$ when DRF=4. Then we reconstruct low-dose PET images using the standard OSEM method \cite{OSEM}. Since the voxel size differs across centers, we uniformly resample the images of different centers so that their voxel size becomes 2$\times$2$\times$2 ${mm}^3$. In the training phase, we unfold images into small patches (uniformly sampling 1024 patches from 20 patients per center) with a shape of 64$\times$64$\times$64. In the testing phase, the whole estimated PET image is acquired by merging patches together. 

 To evaluate the model performance, we choose the PSNR metric for image quantitative evaluation. For clinical evaluation, to address the accuracy of the standard uptake value (SUV) that most radiologists care about, we follow the paper \cite{CycleGAN-mia-2020} to calculate the bias of $SUV_{mean}$ and $SUV_{max}$ (denoted as $B_{mean}$ and $B_{max}$, respectively) between low-dose and full-dose images in lesion regions.

\begin{table}[t] 
\centering
\caption{Results on $C_{kn}$. The \textbf{Best} and the \underline{Second-Best} Results are Highlighted. \\
*: Significant Difference at $p<0.05$ between Comparison Method and Our Method.} 
\resizebox{0.98\textwidth}{!}{
\begin{tabular}{cc|ccccc|ccccc|ccccc}
\toprule
\multicolumn{2}{c|}{\multirow{2}{*}{Methods}} & \multicolumn{5}{c|}{PSNR↑}                                                                               & \multicolumn{5}{c|}{$B_{mean}$↓}                                                                              & \multicolumn{5}{c}{$B_{max}$↓}                                                                                \\ \cline{3-17} 
\multicolumn{2}{c|}{}                         & $C_1$          & $C_2$          & $C_3$          & \multicolumn{1}{c|}{$C_4$}          & Avg.           & $C_1$           & $C_2$           & $C_3$           & \multicolumn{1}{c|}{$C_4$}           & Avg.            & $C_1$           & $C_2$           & $C_3$           & \multicolumn{1}{c|}{$C_4$}           & Avg.            \\ \midrule
\multirow{2}{*}{(i)}        & 3D-cGAN         & 47.30*          & 44.97*          & 45.15*          & \multicolumn{1}{c|}{43.08*}          & 45.13*          & 0.0968*          & $\underline{0.0832}$    & 0.0795*          & \multicolumn{1}{c|}{0.1681*}          & 0.1069*          & 0.1358*          & 0.1696*          & 0.1726*          & \multicolumn{1}{c|}{0.2804*}          & 0.1896*         \\
                            & 3D CVT-GAN      & 47.46*          & 45.17*          & 45.94*          & \multicolumn{1}{c|}{44.04*}          & 45.65*          & $\underline{0.0879}$    & 0.0972*          & 0.0594*          & \multicolumn{1}{c|}{0.1413*}          & 0.0965*          & $\underline{0.1178}$*    & 0.1591*          & 0.1652*          & \multicolumn{1}{c|}{0.2224*}          & 0.1661*          \\ \hline
\multirow{3}{*}{(ii)}       & FedAVG          & 47.43*          & 44.62*          & 45.61*          & \multicolumn{1}{c|}{43.75*}          & 45.35*          & 0.0985*          & 0.0996*          & 0.1006*          & \multicolumn{1}{c|}{0.2202*}          & 0.1122*          & 0.1459*          & $\underline{0.1546}$*    & 0.2011*          & \multicolumn{1}{c|}{0.2663*}          & 0.1920*          \\
                            & FL-MRCM         & 47.81*          & 45.56*          & $\underline{46.10}$*    & \multicolumn{1}{c|}{44.31*}          & 45.95*          & 0.0939*          & 0.0929*          & 0.0631*          & \multicolumn{1}{c|}{0.1344*}          & 0.0961*          & 0.1571*          & 0.1607*          & 0.1307*          & \multicolumn{1}{c|}{0.1518*}          & 0.1501*          \\
                            & FTL             & $\underline{48.05}$*    & $\underline{45.62}$*    & 46.01*          & \multicolumn{1}{c|}{$\underline{44.75}$*}    & $\underline{46.11}$*    & 0.0892          & 0.0945*          & $\underline{0.0587}$*    & \multicolumn{1}{c|}{$\underline{0.0895}$}    & $\underline{0.0830}$*    & 0.1243*          & 0.1588*          & $\underline{0.0893}$    & \multicolumn{1}{c|}{$\underline{0.1436}$}    & $\underline{0.1290}$*    \\ \hline
                        & DRMC            & \textbf{49.48} & \textbf{46.32} & \textbf{46.71} & \multicolumn{1}{c|}{\textbf{45.01}} & \textbf{46.88} & \textbf{0.0844} & \textbf{0.0792} & \textbf{0.0491} & \multicolumn{1}{c|}{\textbf{0.0880}} & \textbf{0.0752} & \textbf{0.1037} & \textbf{0.1313} & \textbf{0.0837} & \multicolumn{1}{c|}{\textbf{0.1431}} & \textbf{0.1155} \\ \bottomrule
\end{tabular}}
\label{tab:methods_comparison}
\end{table}

\begin{table*}[t]
\begin{floatrow}
\capbtabbox{ 
\resizebox{0.41\textwidth}{!}{
\begin{tabular}{cc|cc|cc|cc}
\toprule
\multicolumn{2}{c|}{\multirow{2}{*}{Methods}} & \multicolumn{2}{c|}{PSNR↑}       & \multicolumn{2}{c|}{$B_{mean}$↓} & \multicolumn{2}{c}{$B_{max}$↓}   \\ \cline{3-8} 
\multicolumn{2}{c|}{}                         & $C_5$          & $C_6$          & $C_5$         & $C_6$           & $C_5$         & $C_6$           \\ \midrule 
\multirow{2}{*}{(i)}        & 3D-cGAN         & 26.53*          & 46.07*          & $\textendash$ & 0.1956*          & $\textendash$ & 0.1642*          \\
                            & 3D CVT-GAN      & 27.11*          & 46.03*          & $\textendash$ & $\underline{0.1828}$          & $\textendash$ & 0.1686*          \\ \hline
\multirow{3}{*}{(ii)}       & FedAVG          & 27.09*          & 46.48*          & $\textendash$ & 0.1943*          & $\textendash$ & 0.2291*          \\
                            & FL-MRCM         & 25.38*          & 47.08*          & $\textendash$ & 0.1998*          & $\textendash$ & 0.1762*          \\
                            & FTL             & $\underline{27.38}$*    & $\underline{ 48.05}$*    & $\textendash$ & 0.1898*    & $\textendash$ & $\underline{0.1556}$*    \\ \hline
                        & DRMC            & \textbf{28.54} & \textbf{48.26} & $\textendash$ & \textbf{0.1814} & $\textendash$ & \textbf{0.1483} \\ \bottomrule
\end{tabular}
}
}{
 \caption{Results on $C_{ukn}$.}
 \label{tab:unknown_center_result}
}

\capbtabbox{ 

\resizebox{0.45\textwidth}{!}{ 
\renewcommand{\arraystretch}{1.42}
\begin{tabular}{c|ccc|ccc} 
\toprule 
\multirow{2}{*}{Methods} & \multicolumn{3}{c|}{$C_{kn}$}                      & \multicolumn{3}{c}{$C_{ukn}$}                      \\ \cline{2-7} 
                         & PSNR↑           & $B_{mean}$↓      & $B_{max}$↓       & PSNR↑           & $B_{mean}$↓      & $B_{max}$↓       \\ \midrule
w/o H                      & 46.64*          & 0.0907*          & 0.1436*          & 38.23*          & $\underline{0.1826}$    & 0.1548*          \\
Softmax                    & $\underline{46.70}$*    & $\underline{0.0849}$*    & $\underline{0.1277}$*    & 38.33          & 0.1864*          & $\underline{0.1524}$*    \\
Top-2 Gating                   & 46.61*          & 0.0896*          & 0.1295*          & $\underline{38.38}$    & 0.1867*          & 0.1564*          \\
DRMC                     & \textbf{46.88} & \textbf{0.0752} & \textbf{0.1155} & \textbf{38.40} & \textbf{0.1814} & \textbf{0.1483} \\ \bottomrule
\end{tabular}
}
}{
 \caption{Routing Ablation Results.}
 \label{tab:ablation}
 \small
} 

\end{floatrow} 
\end{table*}

\subsection{Implementation}
Unless specified otherwise, the intermediate channel number, expert number in a bank, and Transformer block number are 64, 3, and 5, respectively. We employ Adam optimizer with a learning rate of $10^{-4}$. We implement our method with Pytorch using a workstation with 4 NVIDIA A100 GPUs with 40GB memory (1 GPU per center). In each training iteration, each GPU independently samples data from a single center. After the loss calculation and the gradient back-propagation, the gradients of different GPUs are then synchronized. We train our model for 200 epochs in total as no significant improvement afterward.

% \subsection{Evaluation} 
% We choose the PSNR metric for image quantitative evaluation. In addition, several studies \cite{CycleGAN-mia-2020} have shown that traditional quantitative assessment metrics (\textit{e.g.} PSNR) may not accurately reflect the accuracy of the standard uptake value (SUV) that most radiologists care about. We thus follow the paper \cite{CycleGAN-mia-2020} to calculate the bias of $SUV_{mean}$ and $SUV_{max}$ (denoted as $B_{mean}$ and $B_{max}$, respectively) between low-dose and full-dose images for clinical evaluation in lesion regions. 

% \subsection{Evaluation} 
% We choose the PSNR metric for image quantitative evaluation. For clinical evaluation, to address the accuracy of the standard uptake value (SUV) that most radiologists care about, we follow the paper \cite{CycleGAN-mia-2020} to calculate the bias of $SUV_{mean}$ and $SUV_{max}$ (denoted as $B_{mean}$ and $B_{max}$, respectively) between low-dose and full-dose images in lesion regions. 

% \begin{equation}
% \begin{aligned}
% B_{mean} = \frac{\left|SUV^{EPET}_{mean}-SUV^{FPET}_{mean}\right|}{SUV^{FPET}_{mean}}, 
% B_{max} = \frac{\left|SUV^{EPET}_{max}-SUV^{FPET}_{max}\right|}{SUV^{FPET}_{max}}, 
% \end{aligned}
% \end{equation} 
% where $B_{mean}$ and $B_{max}$ are bias of $SUV_{mean}$ and $SUV_{max}$, respectively.

% \usepackage{multirow}

\subsection{Comparative Experiments}
We compare our method with five methods of two types. (i) 3D-cGAN \cite{3D-cGAN} and 3D CVT-GAN \cite{3DCVT-GAN} are two state-of-the-art methods for single center PET image synthesis. (ii) FedAVG\cite{FL-MR-FedAVG,FL-MRCM}, FL-MRCM\cite{FL-MRCM}, and FTL\cite{FTL-PET-2022} are three federated learning methods for privacy-preserving multi-center medical image synthesis. All methods are trained using data from $C_{kn}$ and tested over both $C_{kn}$ and $C_{ukn}$. For methods in (i), we regard $C_{kn}$ as a single center and mix all data together for training. For federated learning methods in (ii), we follow the "$\textbf{Mix}$" mode (upper bound of FL-based methods) in the paper \cite{FL-MRCM} to remove the privacy constraint and keep the problem setting consistent with our multi-center study.

\noindent
\textbf{Comparison Results for Known Centers.} As can be seen in Table.~\ref{tab:methods_comparison}, in comparison with the second-best results, DRMC boosts the performance by 0.77 dB PSNR, 0.0078 $B_{mean}$, and 0.0135 $B_{max}$. This is because our DRMC not only leverages shared knowledge by sharing some experts but also preserves center-specific information with the help of the sparse routing strategy. Further evaluation can be found in the \textbf{supplement}.

\begin{table}[t] 
\centering
\caption{Comparison results for Specialized Models and Generalist Models.} 
\resizebox{1\textwidth}{!}{
\begin{tabular}{cc|cl|ccccc|ccccc|ccccc}
\toprule
\multicolumn{2}{c|}{\multirow{3}{*}{Methods}}                & \multicolumn{2}{c|}{\multirow{3}{*}{Train   Centers}} & \multicolumn{5}{c|}{PNSR↑}                                                                                      & \multicolumn{5}{c|}{$B_{mean}$↓}                                                                                    & \multicolumn{5}{c}{$B_{max}$↓}                                                                                      \\ \cline{5-19} 
\multicolumn{2}{c|}{}                                        & \multicolumn{2}{c|}{}                                 & \multicolumn{4}{c|}{Test Centers}                                                      & \multirow{2}{*}{Avg.} & \multicolumn{4}{c|}{Test Centers}                                                          & \multirow{2}{*}{Avg.} & \multicolumn{4}{c|}{Test Centers}                                                          & \multirow{2}{*}{Avg.} \\ \cline{5-8} \cline{10-13} \cline{15-18}
\multicolumn{2}{c|}{}                                        & \multicolumn{2}{c|}{}                                 & $C_1$          & $C_2$          & $C_3$          & \multicolumn{1}{c|}{$C_4$}          &                       & $C_1$           & $C_2$           & $C_3$           & \multicolumn{1}{c|}{$C_4$}           &                       & $C_1$           & $C_2$           & $C_3$           & \multicolumn{1}{c|}{$C_4$}           &                       \\ \midrule
\multicolumn{1}{c|}{}            & \multirow{4}{*}{Baseline} & \multicolumn{2}{c|}{$C_1$}                            & $\underline{48.89}$*    & 45.06*          & 43.94*          & \multicolumn{1}{c|}{41.55*}          & 44.86*                 & $\underline{0.0849}$    & 0.0949*          & 0.1490*           & \multicolumn{1}{c|}{0.2805*}          & 0.1523*                & $\underline{0.1207}$*    & 0.1498*          & 0.3574*          & \multicolumn{1}{c|}{0.4713*}          & 0.2748*                \\
\multicolumn{1}{c|}{Specialized} &                           & \multicolumn{2}{c|}{$C_2$}                            & 47.05*          & $\underline{46.08}$*    & 43.82*          & \multicolumn{1}{c|}{41.53*}          & 44.62*                 & 0.0933*          & \textbf{0.0557}* & 0.1915*          & \multicolumn{1}{c|}{0.2247*}          & 0.1413*                & 0.1326*          & \textbf{0.1243}* & 0.3275*          & \multicolumn{1}{c|}{0.4399*}          & 0.2561*                \\
\multicolumn{1}{c|}{Model}       &                           & \multicolumn{2}{c|}{$C_3$}                            & 44.04*          & 41.00*          & $\underline{46.52}$*    & \multicolumn{1}{c|}{44.07*}          & 44.11*                 & 0.2366*          & 0.2111*          & \textbf{0.0446} & \multicolumn{1}{c|}{0.1364*}          & 0.1572*                & 0.4351*          & 0.5567*          & \textbf{0.0729}* & \multicolumn{1}{c|}{0.1868*}          & 0.3129*                \\
\multicolumn{1}{c|}{}            &                           & \multicolumn{2}{c|}{$C_4$}                            & 44.41*          & 41.39*          & 46.01*          & \multicolumn{1}{c|}{$\underline{44.95}$}    & 44.29*                 & 0.2462*          & 0.2063*          & 0.0897*          & \multicolumn{1}{c|}{$\underline{0.0966}$*}    & 0.1597*                & 0.4887*          & 0.5882*          & 0.1222*          & \multicolumn{1}{c|}{$\underline{0.1562}$*}    & 0.3388*                \\ \hline
\multicolumn{1}{c|}{Generalist}  & Baseline                  & \multicolumn{2}{c|}{$C_1$, $C_2$, $C_3$, $C_4$}       & 47.59*          & 44.73*          & 46.02*          & \multicolumn{1}{c|}{44.20*}          & $\underline{45.64}$*           & 0.0924*          & 0.0839*          & 0.0844*          & \multicolumn{1}{c|}{0.1798*}          & $\underline{0.1101}$*         & 0.1424*          & 0.1424*          & 0.1579*          & \multicolumn{1}{c|}{0.2531*}          & $\underline{0.1740}$*          \\
\multicolumn{1}{c|}{Model}       & DRMC                      & \multicolumn{2}{c|}{$C_1$, $C_2$, $C_3$, $C_4$}       & \textbf{49.48} & \textbf{46.32} & \textbf{46.71} & \multicolumn{1}{c|}{\textbf{45.01}} & \textbf{46.88}        & \textbf{0.0844} & $\underline{0.0792}$    & $\underline{0.0491}$    & \multicolumn{1}{c|}{\textbf{0.0880}} & \textbf{0.0752}       & \textbf{0.1037} & $\underline{0.1313}$    & $\underline{0.0837}$    & \multicolumn{1}{c|}{\textbf{0.1431}} & \textbf{0.1155}       \\ \bottomrule
\end{tabular}
}
\label{tab:spe_vs_gen}
\end{table} 

\noindent
\textbf{Comparison Results for Unknown Centers.} We also test the model generalization ability to unknown centers $C_5$ and $C_6$. $C_5$ consists of normal brain data (without lesion) that is challenging for generalization. As the brain region only occupies a small portion of the whole-body data in the training dataset but has more sophisticated structure information. $C_6$ is a similar center to $C_1$ but has different working locations and imaging preferences. The quantitative results are shown in Table.~\ref{tab:unknown_center_result} and the visual results are shown in Fig.~\ref{fig:center_interference} (a). DRMC achieves the best results by dynamically utilizing existing experts' knowledge for generalization. On the contrary, most comparison methods process data in a static pattern and unavoidably produce mishandling of out-of-distribution data.

Furthermore, we evaluate the performance of different models on various DRF data on $C_6$, and the results are available in the \textbf{supplement}. These results indicate that our method demonstrates strong robustness.

\begin{figure*}[t]
\centering 

\subfigure[Visual comparison on the unknown center $C_5$]{
    \includegraphics[height=1in]{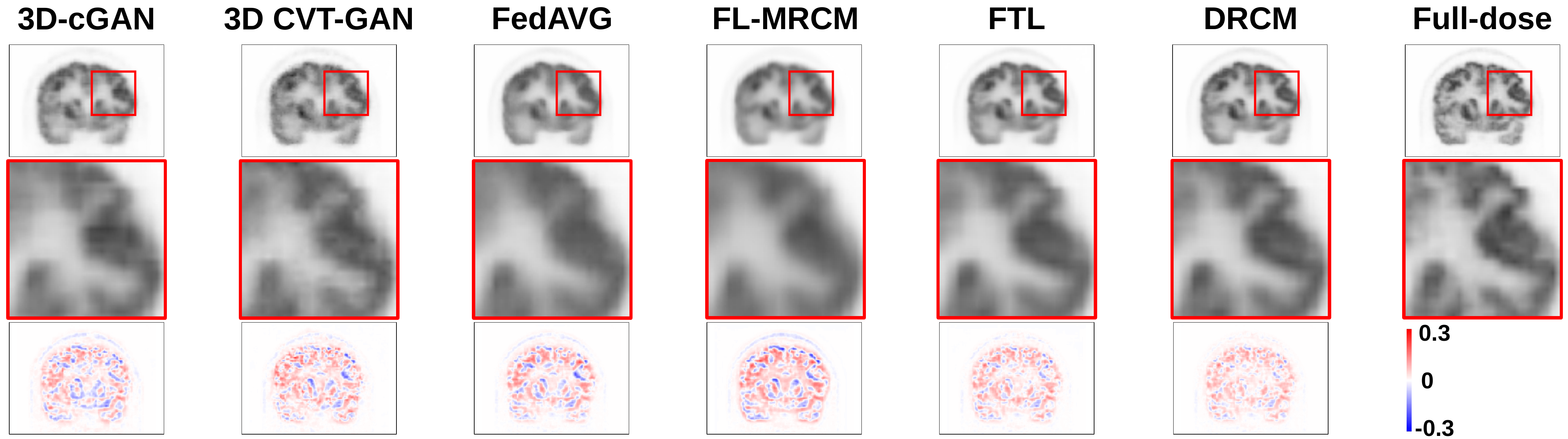}

} 
\\

\subfigure[Top-1 Expert]{
    \includegraphics[height=.6in]{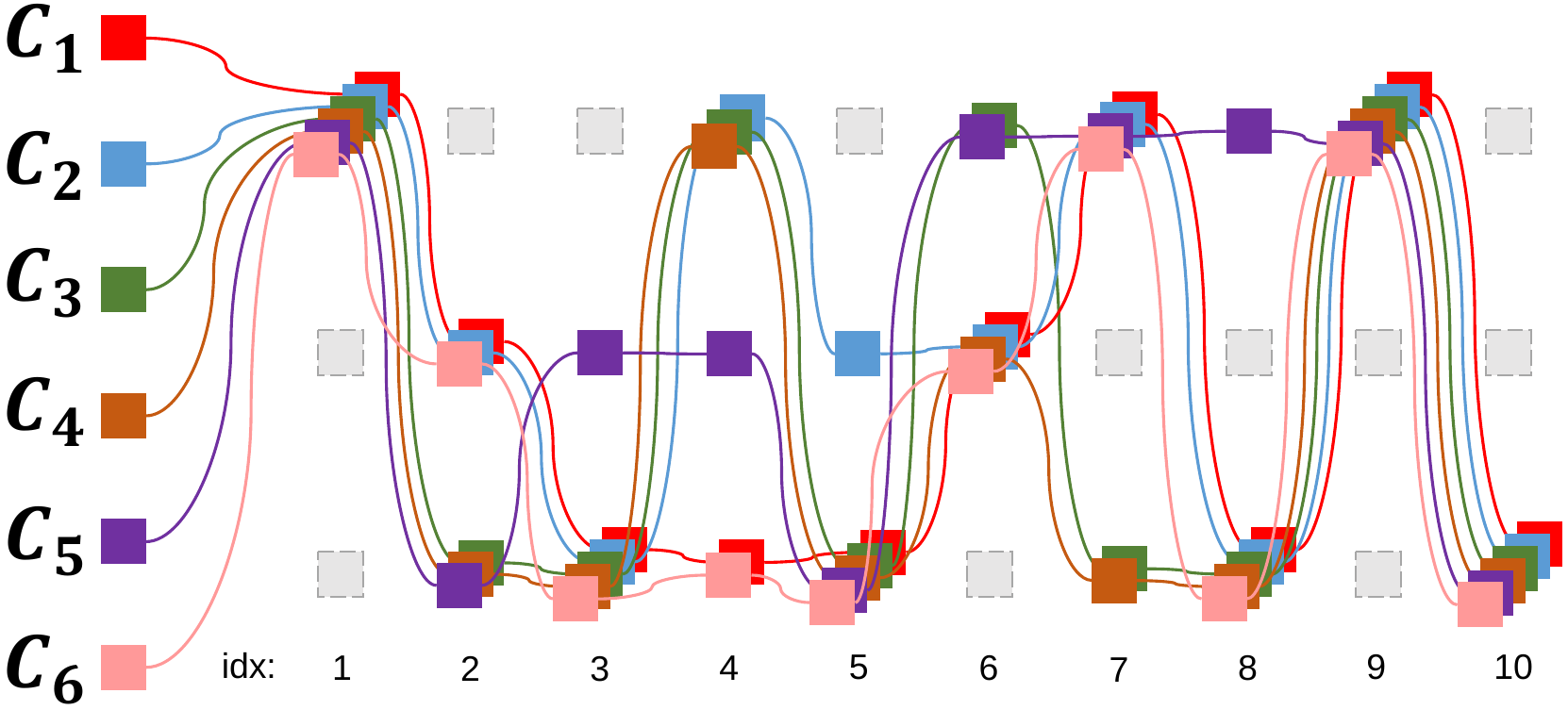}

}
\subfigure[PSNR/N]{
    \includegraphics[height=.6in]{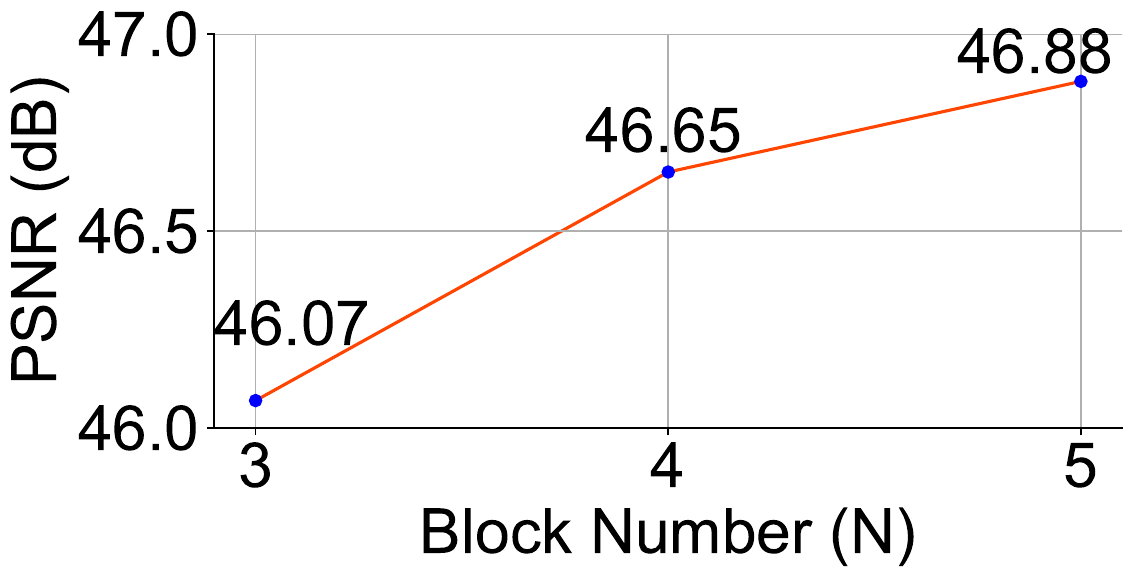}

}
\subfigure[PSNR/M]{
    \includegraphics[height=.6in]{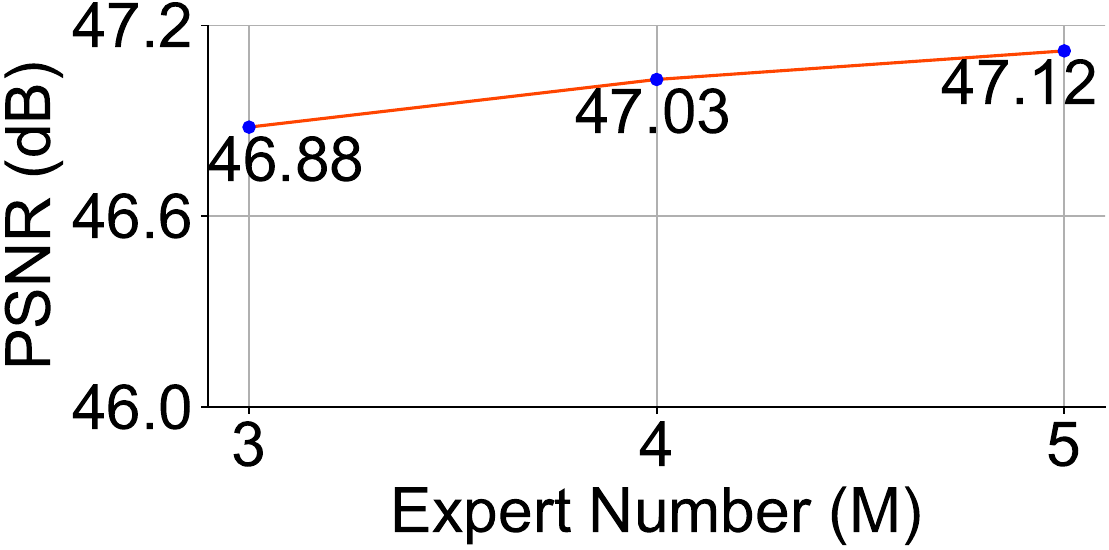}

}

\caption{Figures of different experiments.}
\label{fig:fig_of_different_experiments}
\end{figure*}

\subsection{Ablation Study}
\textbf{Specialized Model vs. Generalist Model.} As can be seen in Table.~\ref{tab:spe_vs_gen}, the baseline model (using 15 base blocks) individually trained for each center acquires good performance on its source center. But it suffers performance drop on other centers. The baseline model trained over multiple centers greatly enhances the overall results. But due to the center interference issue, its performance on a specific center is still far from the corresponding specialized model. DRMC mitigates the interference with dynamic routing and achieves comparable performance to the specialized model of each center.

\noindent
\textbf{Ablation Study of Routing Strategy.} To investigate the roles of major components in our routing strategy, we conduct ablation studies through (i) removing the condition of hidden representation $H$ that builds cross-layer connection, and replacing ReLU activation with (ii) softmax activation \cite{domain_attention-2019} and (iii) top-2 gating \cite{moe-2017}. The results are shown in Table.~\ref{tab:ablation}. We also analyze the interpretability of the routing by showing the distribution of different layers' top-1 weighted experts using the testing data. As shown in Fig.~\ref{fig:fig_of_different_experiments} (b), different centers show similarities and differences in the expert distribution. For example, $C_6$ shows the same distribution with $C_1$ as their data show many similarities, while $C_5$ presents a very unique way since brain data differs a lot from whole-body data.

\noindent
\textbf{Ablation Study of Hyperparameters.} In Fig.~\ref{fig:fig_of_different_experiments} (c) and (d), we show ablation results on  expert number ($M$) and block number ($N$). We set $M$=3 and $N$=5, as this configuration has demonstrated good performance while maintaining acceptable computational complexity.

\section{Conclusion}
In this paper, we innovatively propose a generalist model with dynamic routing (DRMC) for multi-center PET image synthesis. To address the center interference issue, DRMC sparsely routes data from different centers to different experts. Experiments show that DRMC achieves excellent generalizability. 
% For citations of references, we prefer the use of square brackets
% and consecutive numbers. Citations using labels or the author/year
% convention are also acceptable. The following bibliography provides
% a sample reference list with entries for journal
% articles~\cite{ref_article1}, an LNCS chapter~\cite{ref_lncs1}, a
% book~\cite{ref_book1}, proceedings without editors~\cite{ref_proc1},
% and a homepage~\cite{ref_url1}. Multiple citations are grouped
% \cite{ref_article1,ref_lncs1,ref_book1},
% \cite{ref_article1,ref_book1,ref_proc1,ref_url1}.
% \clearpage
\bibliographystyle{splncs04}
\bibliography{ref} 

\newpage

\section*{Supplement}
\textbf{Center Interference.} To quantify the interference of the $j$-th center task on the $i$-th center task, we estimate the change in loss $\mathcal{L}_i$ for the $i$-th center task when optimizing the shared parameters $\theta$ according to the $j$-th center task's loss $\mathcal{L}_j$ as follows:

\begin{equation} 
\begin{aligned}
\Delta_j \mathcal{L}_{i}\left(X_i\right) &\doteq \mathbb{E}_{X_j}\left(\mathcal{L}_{i}\left(X_i ; \theta\right)-\mathcal{L}_{i}(X_i ; \theta-\lambda \frac{\nabla_\theta \mathcal{L}_{j}\left(X_j\right)}{\left\|\nabla_\theta \mathcal{L}_{j}\left(X_j\right)\right\|})\right), \\
&\approx \lambda \mathbb{E}_{X_j}\left({\frac{\nabla_\theta \mathcal{L}_{j}\left(X_j\right)}{\left\|\nabla_\theta \mathcal{L}_{j}\left(X_j\right)\right\|}}^T \nabla_\theta \mathcal{L}_{i}\left(X_i\right)\right), 
\end{aligned}
\end{equation} 

where $X_i$ and $X_j$ are the sampled training batches of the $i$-th and $j$-th centers, respectively. In the implementation, we sample 100 batches from each center for interference calculation. The interference of the $j$-th center task on the $i$-th center task can then be quantified as follows:
\begin{equation}
\mathcal{I}_{i, j}=\mathbb{E}_{X_i}\left(\frac{\Delta_j \mathcal{L}_{i}\left(X_i\right)}{\Delta_i \mathcal{L}_{i}\left(X_i\right)}\right),
\end{equation}
where the denominator is utilized to normalize the scale of the loss change.

\noindent
\textbf{SSIM Evaluation.} To further assess the performance of our method, we compare the SSIM metric between our method and other comparison methods. The results are presented in Table.~\ref{tab:ssim}. The results indicate that DRMC achieves the highest performance on the SSIM metric.

\begin{table}[h]
\centering
\caption{SSIM Results on $C_{kn}$. The \textbf{Best} and the \underline{Second-Best} Results are Highlighted. \\
*: Significant Difference at $p<0.05$ between Comparison Method and Our Method.} 
\begin{tabular}{cc|ccccc}
\hline
\multicolumn{2}{c|}{\multirow{2}{*}{Methods}} & \multicolumn{5}{c}{SSIM}                                                                                       \\ \cline{3-7} 
\multicolumn{2}{c|}{}                         & $C_1$           & $C_2$           & $C_3$           & \multicolumn{1}{c|}{$C_4$}            & Avg.             \\ \hline
\multirow{2}{*}{(i)}        & 3D-cGAN         & 0.8524*         & 0.8267*         & 0.8077*         & \multicolumn{1}{c|}{0.7972*}          & 0.8210*          \\
                            & 3D CVT-GAN      & 0.8837*         & 0.8931*         & \underline{0.9198}*         & \multicolumn{1}{c|}{0.8770*}          & 0.8934*          \\ \hline
\multirow{3}{*}{(ii)}       & FedAVG          & 0.8997*         & \underline{0.9091}*         & 0.9188*         & \multicolumn{1}{c|}{0.8802*}          & \underline{ 0.9020}*    \\
                            & FL-MRCM         & 0.8960*         & 0.8812*         & 0.8777*         & \multicolumn{1}{c|}{0.8482*}          & 0.8758*          \\
                            & FTL             & \underline{ 0.9045}*   & 0.8889*   & 0.9056*  & \multicolumn{1}{c|}{\underline{ 0.8816}*}    & 0.8952*          \\ \hline
                            & DRMC            & \textbf{0.9173} & \textbf{0.9172} & \textbf{0.9284} & \multicolumn{1}{c|}{\textbf{0.9010}} & \textbf{0.9160} \\ \hline
\end{tabular} 
\label{tab:ssim}
\end{table}

\noindent
\textbf{Evaluation on Different DRF Data.} 
To verify the robustness of the model on different dosage data, we conducted tests on the unknown center $C_6$. Table~\ref{tab:drf} presents the comparison results, demonstrating that our DRMC exhibits superior generalizability across different DRF data.

% \usepackage{multirow}
% \usepackage[normalem]{ulem}
\begin{table}[t]
\centering 
\caption{Testing Results on Different DRF Data from $C_6$. The \textbf{Best} and the \underline{Second-Best} Results are Highlighted. *: Significant Difference at $p<0.05$ between Comparison Method and Our Method.} 
\resizebox{0.98\textwidth}{!}{
\begin{tabular}{cc|ccccc|ccccc|ccccc}
\hline
\multicolumn{2}{c|}{\multirow{2}{*}{Methods}} & \multicolumn{5}{c|}{PSNR}                                                                               & \multicolumn{5}{c|}{$B_{mean}$}                                                                              & \multicolumn{5}{c}{$B_{max}$}                                                                                \\ \cline{3-17} 
\multicolumn{2}{c|}{}                         & DRF=3          & DRF=4          & DRF=6          & \multicolumn{1}{c|}{DRF=12}         & Avg.           & DRF=3           & DRF=4           & DRF=6           & \multicolumn{1}{c|}{DRF=12}          & Avg.            & DRF=3           & DRF=4           & DRF=6           & \multicolumn{1}{c|}{DRF=12}          & Avg.            \\ \hline
\multirow{2}{*}{(i)}        & 3D-cGAN         & 48.79*         & 48.63*         & 47.86*         & \multicolumn{1}{c|}{46.07*}         & 47.84*         & 0.0881*         & 0.0886*         & 0.1024*         & \multicolumn{1}{c|}{0.1956*}         & 0.1187          & 0.0653*         & 0.0744*         & 0.1086*         & \multicolumn{1}{c|}{0.1642*}         & 0.1031*         \\
                            & 3D CVT-GAN      & 48.85*         & 48.57*         & 47.87*         & \multicolumn{1}{c|}{46.03*}         & 47.83*         & 0.0951*         & 0.1028*         & 0.1191*         & \multicolumn{1}{c|}{\underline{0.1828}*}   & 0.1250          & 0.0677*         & 0.0821*         & 0.1032*         & \multicolumn{1}{c|}{0.1686*}         & 0.1054*         \\ \hline
\multirow{3}{*}{(ii)}       & FedAVG          & 48.17*         & 48.23*         & 47.89*         & \multicolumn{1}{c|}{46.48*}         & 47.69*         & 0.1112*         & 0.1258*         & 0.1303*         & \multicolumn{1}{c|}{0.1943*}         & 0.1404          & 0.0624*         & 0.0781*         & 0.1021*         & \multicolumn{1}{c|}{0.2291*}         & 0.1179*         \\
                            & FL-MRCM         & 50.49*         & 50.21*         & 48.65*         & \multicolumn{1}{c|}{47.08*}         & 49.11*         & 0.1012*         & 0.1038*         & 0.1042*         & \multicolumn{1}{c|}{0.1998*}         & 0.1273          & 0.0543*         & 0.0956*         & 0.1041*         & \multicolumn{1}{c|}{0.1762*}         & 0.1076*         \\
                            & FTL             & \underline{51.01}    & \underline{50.89}    & \underline{49.43}*   & \multicolumn{1}{c|}{\underline{48.05}*}   & \underline{49.85}*   & \underline{0.0553}*   & \underline{0.0878}    & \underline{0.0986}*    & \multicolumn{1}{c|}{0.1898*}          & \underline{0.1079}*    & \underline{0.0484}    & \underline{0.0524}*   & \underline{0.0724}*   & \multicolumn{1}{c|}{\underline{0.1556}*}   & \underline{0.0822}*   \\ \hline
(iii)                       & DRMC            & \textbf{51.04} & \textbf{50.95} & \textbf{49.59} & \multicolumn{1}{c|}{\textbf{48.26}} & \textbf{49.96} & \textbf{0.0438} & \textbf{0.0836} & \textbf{0.0929} & \multicolumn{1}{c|}{\textbf{0.1814}} & \textbf{0.1004} & \textbf{0.0455} & \textbf{0.0457} & \textbf{0.0655} & \multicolumn{1}{c|}{\textbf{0.1483}} & \textbf{0.0763} \\ \hline
\end{tabular}}
\label{tab:drf}
\end{table}

\end{document}